\documentclass[useAMS,usenatbib]{mn2e}
\usepackage{graphicx}
\usepackage{natbib}
\usepackage{amssymb}
\usepackage{setspace}
\usepackage{amssymb}
\usepackage{epsfig,color}
\usepackage{multicol}

\long\def\symbolfootnote[#1]#2{\begingroup%
\def\thefootnote{\fnsymbol{footnote}}\footnote[#1]{#2}\endgroup} 

\newcommand{\Msol}{~$M_{\odot}$} 
\newcommand{\cc}{~cm$^{-3}$}   
\newcommand{\wkinii}{\texttt{wkin2}}
\newcommand{\wkiniv}{\texttt{wkin4}}
\newcommand{\wkinth}{\texttt{wkin2th2}}
\newcommand{\wkinmag}{\texttt{wkin2mag}}

\newcommand{\tvelweCO}{$^{12}$CO~}  
\newcommand{\thirthCO}{$^{13}$CO~}  
\newcommand{\CeightO}{C$^{18}$O~}
\newcommand{\HCO}{H$^{13}$CO$^+$~($J=1-0$)~}
\newcommand{\CS}{CS~($J=1-0$)~}
\title[Clump mass function and MC evolution: II.]{Clump mass function at an early stage of molecular cloud evolution: II. Galactic cloud complexes}
\author[Veltchev, Donkov \& Klessen]
  {Todor V.~Veltchev$^{1,2\,\star}$, Sava Donkov$^3$, and Ralf S.~Klessen$^2$ \\
  $^1$University of Sofia, Faculty of Physics, 5 James Bourchier Blvd., 1164 Sofia, Bulgaria\\
  $^2$Universit\"at Heidelberg, Zentrum f\"ur Astronomie, Institut f\"ur Theoretische Astrophysik, Albert-\"Uberle-Str. 2, 69120 Heidelberg, Germany\\
  $^3$Department of Applied Physics, Technical University, 8 Kliment Ohridski Blvd., 1000 Sofia, Bulgaria}

\date{Accepted 2013, April 22}

\pagerange{\pageref{firstpage}--\pageref{lastpage}} \pubyear{2012}

\begin{document}

\label{firstpage}

\maketitle

\begin{abstract}
The statistical approach for derivation of the clump mass function (ClMF) developed by Donkov, Veltchev \& Klessen is put to observational test through comparison with mass distributions of clumps from molecular emission and dust continuum maps of Galactic cloud complexes, obtained by various authors. The results indicate gravitational boundedness of the dominant clump population, with or without taking into account the contribution of their thermal and magnetic energy. The ClMF can be presented by combination of two power-law functions separated by a characteristic mass from about ten to hundreds solar masses. The slope of the intermediate-mass ClMF is shallow and nearly constant ($-0.25\gtrsim\Gamma_{\rm IM}\gtrsim-0.55$) while the high-mass part is fitted by models that imply gravitationally unstable clumps and exhibit slopes in a broader range ($-0.9\gtrsim\Gamma_{\rm IM}\gtrsim-1.6$), centered at the value of the stellar initial mass function ($\Gamma_{\rm HM}\gtreqless-1.3$).
\end{abstract}


\section{Introduction} 
Dense clumps in molecular clouds (MCs) are typical sites of star formation as they are often associated with young stellar objects. Their origin can be sought in the early epoch of cloud evolution when supersonic turbulence creates sets of condensations in the cold, mainly molecular gas. Recent numerical simulations indicate their mean densities, sizes and masses vary in ranges $10^2-10^4$\cc, $0.04 - 1$~pc and $0.1-10^3$\Msol, respectively \citep{VS_ea_07, Baner_ea_09, Shet_ea_10}, in consistency with extensive observational data about MC clumps \citep[e.g.,][]{BT_07}. Clump morphology is also diverse: from filamentary to compact, quasi-spherical shapes \citep{Henne_ea_08}. Gravitational stability analysis shows that some clumps are subject to further contraction and collapse and eventually give birth to single stars or stellar clusters. 
\symbolfootnote[0]{$\star$~E-mail: eirene@phys.uni-sofia.bg}

Numerous individual clumps were initially identified on maps of molecular line emissions which trace different density regimes in MCs: \tvelweCO ($n\sim 10^2$\cc), \thirthCO and \CeightO ($n\sim 10^3$\cc), \CS and \HCO ({$n\gtrsim 10^4$\cc}). Such surveys were performed in nearby (at distances $<500$~pc) Galactic cloud complexes like Orion A \citep{Tatematsu_ea_93}, Orion B \citep{Kramer_ea_98}, Taurus \citep{Oni_ea_96}, Ophiuchus \citep{Tachi_ea_00}, Lupus \citep{Hara_ea_99} and many others. Some of these results on the physical parameters of clumps were included in the statistical study of \citet{Tachi_ea_02} who considered a sample of 9 Galactic star-forming regions. In the last decade, further intensive research of MCs by use of some high-density tracers like \HCO \citep{Oni_ea_02, ISK_07, IKS_09} as well of dust continuum \citep{John_ea_01, Kerton_ea_01, JMM_06, RW_05, RW_06a, DiFran_ea_10} and dust extinction \citep[e.g.][]{ALL_07} observations allowed for more precise mapping of cloud structure. Some clumps originally found on emission maps were further decomposed and more compact (typical sizes $\lesssim 0.2$~pc), very dense ($n\gtrsim 10^5$\cc) and probably collapsing clumps were delineated. Some authors call such objects `dense cores'; hereafter, we label them simply {\it cores}. 

It is suggested that cores eventually form stars \citep{BT_07} and thus the study of their mass function (CMF) will enable a better understanding of the physical origin of the stellar initial mass function (IMF) and its possible variations. Indeed, numerous dust continuum and dust extinction observations demonstrate that the CMF resembles the IMF in its shape when fitted by a single power-law \citep{TS_98, John_ea_01}, a combination of two power-law \citep{MAN_98, MA_01, JMM_06, NW-T_07} or a lognormal function \citep{Stanke_ea_06, Enoch_ea_08, Konyves_ea_10}. It therefore has been proposed that the IMF is a direct product of the CMF and a uniform star formation efficiency \citep{ALL_07}. 

Yet it is still unclear how the CMF originates from the mass distribution of the initially formed MC clumps. The clump mass function (ClMF), as derived from molecular line emission surveys of Galactic star-forming regions using low- and moderate-density tracers, varies in shape and slope and often differs substantially from the IMF. Most of the earlier studies of nearby complexes from CO mapping resulted in a single power-law ClMF of a rather shallow slope $-0.3\gtrsim\Gamma\gtrsim-0.8$ without any characteristic mass $M_{\rm ch}$ \citep{SG_90, Blitz_93, WBS_95, Heit_ea_98} -- in contrast to the Salpeter slope $-1.3$ \citep{Sal55} of the high-mass IMF over $M_{\rm ch}\sim 0.5 M_\odot$. Moreover, the ClMF slope $\Gamma$ was found to remain nearly constant for a large range of clump masses $10^{-3}\le m/M_\odot \le10^4$, in different MCs \citep{Kramer_ea_98}. 

Later works allowed for a more detailed mapping of MCs. Emission from CO molecules was found to trace lower density cloud regions. Tracers like C$^{18}$O revealed structures with $n \sim 10^4$\cc~but essentially larger ($l\sim0.1-0.5$~pc) than prestellar cores~\citep{Hara_ea_99, Tachi_ea_00, Tachi_ea_02}. Such compact clumps encompass about 10\% of the cloud mass. It was demonstrated that the derived ClMFs could be represented by a combination of 2 power-law functions as $M_{\rm ch}\sim15-20$\Msol~separates intermediate-mass from high-mass ClMF, with slopes $-0.3\gtrsim\Gamma_{\rm IM}\gtrsim-0.7$ and $\Gamma_{\rm HM}\lesssim-1.5$, respectively. This general picture was confirmed from submillimeter continuum studies of MC complexes at kpc-scale distances. Using $450~\mu m$ and $850~\mu m$ SCUBA maps, Reid \& Wilson derived 2 power-law ClMFs for two massive star-forming regions. For NGC\,7538, they obtained $M_{\rm ch}\sim 100 M_\odot$ and $\Gamma_{\rm IM}\sim0$ while the variations of the high-mass slope seem 
to be more sensitive to the wavelength and close to the Salpeter value: $-1\gtrsim\Gamma_{\rm HM}\gtrsim-1.6$ \citep{RW_05}. For M\,17, the characteristic mass turned out to be an order of magnitude less ($M_{\rm ch}\sim 8-20~M_\odot$), the intermediate mass slopes are positive ($\Gamma_{\rm IM}\sim0.5-1$) and the high-mass ones are  about the value for CO clumps $-0.5\gtrsim\Gamma_{\rm HM}\gtrsim-0.9$ \citep{RW_06a}. Extending further their study to 11 low- and high-mass star-forming regions, those authors conclude that a single power-law ClMF is clearly ruled out and argue for a two power-law function with $-0.2\gtrsim\Gamma_{\rm IM}\gtrsim-0.7$ and $\Gamma_{\rm HM}$ about the Salpeter slope, regardless of the diversity of characteristic masses \citep{RW_06b}.

The predictions of our statistical model \citep[][hereafter, Paper I]{DVK_12} are generally consistent with a two power-law ClMF with shallow intermediate-mass slope $\Gamma_{\rm IM}$ and steeper high-mass slope $\Gamma_{\rm HM}$, allowing for a variety of $M_{\rm ch}$ within two orders of magnitude. In this work they are compared with results on the observational ClMF: i) from molecular-line studies of three nearby star-forming regions: Orion A \citep{Tatematsu_ea_93}, Orion B \citep{Kramer_ea_98}, Taurus \citep{Oni_ea_96}, and of a sample of 9 regions \citep{Tachi_ea_02}; ii) from dust continuum surveys {\it and} CO mappings of 2 kpc-away regions: M\,17 \citep{RW_06a, SG_90} and Rosette \citep{DiFran_ea_10, WBS_95}. We define {\it clumps} as condensations formed through a turbulent cascade during the early MC evolution. The ensemble of clumps generated at given spatial scale $L$ obeys a power-law relationship between clump masses and densities $n\propto m^x$ where the exponent $x=x(L)$ is calculated considering equipartition relations between various forms of energy: gravitational, turbulent (kinetic), thermal (internal) and magnetic \citep[][hereafter, DVK11]{DVK_11}. The ClMF is derived as a superposition of clump mass distributions over a range of scales, taking into account the fractal structure of the cloud (Paper I). 

The physical basis of the model, its free parameters and the construction of the ClMF are recalled in Section \ref{ClMF_model}. The clump mass distribution derived from modeled structure of each considered individual cloud as well as different distributions that fit the composite observational ClMF are presented and commented in Section~\ref{Results}. Section~\ref{Discussion} contains a discussion on the applicability and the restrictions of our approach to predict the observational ClMF and envisions its possible extensions. Our conclusions are summarized in Section~\ref{Summary}.

\section{Structure and clump mass function of individual cloud}  \label{ClMF_model} 

\subsection{Physical framework of the model} \label{Physical_framework}
Our statistical model for clump description is presented in details in DVK11 (Sect. 2 and 3) and Paper I (Sect. 2). Its main assumptions can be summarized as follows:
\begin{enumerate}
\item {\it Scaling laws within the turbulent cloud:} We consider fully developed supersonic turbulence that implies homogeneous and isotropic stochastic medium with a fractal structure and well-defined scaling laws of turbulent velocity, mean density and mean magnetic field. Turbulent flows create density structures at any scale in the inertial range $L_{\rm max}\gtrsim L \gtrsim L_{\rm min}$ through a cascade possibly driven by the very process of cloud formation \citep{KH_10}. We estimate the upper limit $L_{\rm max}=20$~pc adopting a typical size of giant MC $\sim50$~pc as an injection scale and taking in view that the largest scale of the turbulence inertial range is about a factor of 3 less \citep{Kr_ea_07, Pad_ea_06}. The lower limit $L_{\rm min}$ is imposed above the actual end of the inertial range from the construction of our model: the sizes $l$ of all clumps generated at given scale $L$ must be within the inertial range and are typically an order of magnitude less than $L$. Various 
observations show that the inertial range spans at least 3 orders of magnitude, i.e. the size of the smallest generated object $l_{\rm min}$ must be about $0.02$~pc. We consider mainly molecular and isothermal gas with temperatures $T=10-20$~K. Requiring supersonic medium at all fractal scales and by use of a typical velocity scaling (see equation \ref{eq_Larson_1} below), one obtains $l_{\rm min}\gtrsim0.05$~pc which yields a scale of its generation $L_{\rm min}\simeq0.5$~pc. 
 
Turbulent velocity dispersion $u$ and mean mass density $\langle\rho\rangle$ are assumed to scale according to ``Larson's first and second laws'' \citep{Lars_81} whereas the scaling relation of the mean magnetic field $B$ is obtained from its observationally verified relation to the mean mass density \citep[$B\propto\langle \rho \rangle^{1/2}$:][]{Crutch_99}:
\begin{equation}
\label{eq_Larson_1}
 u = 1.1\,L^\beta~~~\rm [km\,s^{-1}]~, 
\end{equation}
\begin{equation}
\label{eq_Larson_2}
 \langle \rho \rangle = 13.6\times10^{-21}\,L^\alpha~~~\rm [g\,cm^{-3}]~, 
\end{equation}
\begin{equation}
\label{eq_mag_scaling}
B = 50\,L^{0.5\alpha}~~~\rm [ \mu G]~,
\end{equation}
A fixed velocity scaling index $\beta$ is chosen. While observational and numerical studies indicate a broad range $0.2\lesssim\beta\lesssim0.65$, the results in this Paper narrow it to $0.2\lesssim\beta\le0.47$ (`soft' velocity scaling). The density scaling index $\alpha=\alpha(L)$ is derived self-similarly from the assumption of mass-density relationship for clumps generated at a given scale $L$ (see below).

\item {\it Lognormal clump density distribution:} Such volumetric distribution of mass density $\rho$ is testified from numerous numerical simulations of supersonic turbulence and is described through a standard lognormal probability density function (pdf): 
\begin{equation}
\label{eq_pdf}
p(s)\,ds=\frac{1}{\sqrt{2\pi \sigma^2}}\,\exp{\Bigg[-\frac{1}{2}\bigg( \frac{s -s_{\rm max}}{\sigma}\bigg)^2 \Bigg]}\,ds~,
\end{equation}
where $s$ is the log density and $(s_{\rm max},~\sigma)$ are the distribution peak and the standard deviation, respectively. In our model, this density statistics is used as the basis of clump statistics as follows. Let $N^{\rm p}_{\rm tot}$ is the total number of pixels in a cloud map and $N^{\rm p}_{\rho}$ is the number of those with density $\rho$. Then 
$$\frac{N^{\rm p}_{\rho}}{N^{\rm p}_{\rm tot}}\sim \frac{1}{\sqrt{2\pi \sigma^2}}\,\exp{\Bigg[-\frac{1}{2}\bigg( \frac{s -s_{\rm max}}{\sigma}\bigg)^2 \Bigg]}~.$$

Since turbulence is assumed to be homogeneous and isotropic and gravity is a central force, one can consider the generated clumps as homogeneous spheres, characterized solely by their size (diameter) $l$. Of course, the real fragments in a turbulent MC with high Mach numbers ($\geq{3}$) delineated by isodensity contours $[\rho-\Delta\rho,~\rho+\Delta\rho]$ can be with complex shapes and even not necessarily connected regions. The essence of our statistical approach is to describe them by an ensemble of $N^{\rm c}_{\rho}\propto (N^{\rm p}_{\rho}/N^{\rm p}_{\rm tot})$ spherical homogeneous clumps with density $\rho$. Thus the `average clump ensemble' generated at given spatial scale $L$ follows a lognormal density distribution like the pdf at that scale.

The parameters of the density distribution: 
\begin{equation}
\label{eq_sigma_PDF}
\sigma^2={\rm ln}\,(1+b^2\,{\cal M}^2)~,~~~s_{\rm max}=-\frac{\sigma^2}{2}
\end{equation}
depend on the spatial scale through the sonic Mach number ${\cal M}=u(L)/c_{\rm s}$ where $c_{\rm s}$ is the sound speed. The turbulence forcing parameter $b$ spans values between $0.33$, for purely solenoidal forcing, and $1.00$, for purely compressive forcing (see \citet{Fed_ea_10}). The modeled ClMFs presented in this Paper favor mainly solenoidal forcing, mixed in a natural way with compressive modes: $b\le 0.46$. A comment on that is included in the Discussion.

\item{\it Clump mass-density-size relationship:} Masses, densities and sizes of clumps in an `average ensemble' are assumed to obey the statistical relationships:
\begin{equation}
\label{eq_n-m}
\ln\Big(\frac{\rho}{\rho_0}\Big)=x\,\ln\Big(\frac{m}{m_0}\Big)
\end{equation}
\begin{equation}
\label{eq_n-l}
\ln\Big(\frac{\rho}{\rho_0}\Big)=\frac{3x}{1-x}\,\ln\Big(\frac{l}{l_0}\Big)~,
\end{equation}
with a choice of normalization units:
\begin{eqnarray}
 \label{eq_norm_rho}
 \rho_0 & \equiv & \langle\rho\rangle \\
 \label{eq_norm_L}
 l_0 & \equiv & \kappa L \\
 \label{eq_norm_m}
 \frac{\rho_0 l_0^3}{m_0} & \propto & \exp\Big(\sigma^2\times\frac{1-x}{x}\Big)
\end{eqnarray}
The dimensionless parameter $\kappa$ accounts for the precision of the clump size `measurement' and can be interpreted as mapping resolution of the scale of clump generation. It is appropriate to set it as a small constant of order of several percent since $l$ is typically an order of magnitude less than $L$. 

According to our turbulent scenario of clump formation, the clump density-size relationship (equation \ref{eq_n-l}) is taken as a self-similar extension of the density scaling law  (equation~\ref{eq_Larson_2}) and hence:
\begin{equation}
 \label{eq_dens_scaling}
 \alpha=\frac{3x}{1-x}~.
\end{equation}

\item{\it Equipartitions between clump energies:} 
A general type of equipartition relation between gravitational $W$ and kinetic $E_{\rm kin}$ energy per unit volume with some coefficient $f_{\rm gk}$ of proportionality
$$|W|\sim f_{\rm gk}E_{\rm kin}~, $$
is expected to hold for structures shaped by turbulence in which gravity gradually takes over; e.g. in regions where turbulence decays locally or where it accumulates material reaching a state of local gravitational instability. A fiducial range $1\le f_{\rm gk}\le 4$ is testified from simulations of the early stage of the clump evolution, before stars have been formed \citep{VS_ea_07}. The analysis of MHD simulations of cloud formation at Galactic scale \citep[length unit $\sim$1~kpc;][]{Pass_ea_95} showed that the considered equipartition can include contributions of thermal (internal) $E_{\rm th}$ and magnetic $E_{\rm mag}$ energy as well \citep{BP_VS_95}.

In our model, we assume such equipartition relations to hold for the `average clump ensemble' and use them to derive the mass-density scaling index $x(L)$ at each scale:
\begin{itemize}
 \item \wkinii~or \wkiniv, equipartition of the gravitational vs. kinetic energy:
 \begin{eqnarray}
 \label{eq_wkin}  
 |W|\sim 2E_{\rm kin}~,\\
 |W|\sim 4E_{\rm kin}~.
 \end{eqnarray}
 
 \item \wkinmag, equipartition of the gravitational vs. kinetic and magnetic energy:
 \begin{equation}
  \label{eq_wkin2mag}  
  |W|\sim 2E_{\rm kin}+E_{\rm mag}~.
 \end{equation}

 \item \wkinth, equipartition of the gravitational vs. kinetic and thermal energy:
 \begin{equation}
  \label{eq_wkin2th2}  
  |W|\sim 2E_{\rm kin}+2E_{\rm th}~.
 \end{equation}
The corresponding equations for the typical member of the `average clump ensemble' which are used to derive $x(L)$ are listed in Paper I (Appendix A and B). 
\end{itemize}

\end{enumerate}

\subsection{Free parameters of the model and cloud structure from extinction maps}  \label{Cloud_structure}
In our model, a MC is considered as a hierarchical set of spatial scales and its structure is described through the solutions $x(L)$ obtained for a chosen equipartition relation. They depend on four free parameters: velocity-scaling index $\beta$, turbulent forcing parameter $b$, mapping resolution $\kappa$ (equation \ref{eq_norm_L}) and temperature $T$. The parameters' ranges of variation that yield plausible solutions can be restricted by use of observational studies of MC structure. The  work of \citet[][hereafter, LAL10]{LAL_10}, based on dust extinction maps of Galactic cloud complexes, corresponds best to our approach. These authors define an `effective radius $R_{\rm s}=\sqrt{S/\pi}$ of a subregion or a set of subregions with total area $S$ which we interpret as an observational counterpart of our notion of `spatial scale', i.e. $L=2R_{\rm s}$. Then the MC structure is described by the relationship between the effective radius and the mass within the total area $S$: $M_{\rm s} \propto R_{\rm s}^{\gamma}$ (see Fig. 
2 in LAL10). On the other hand, the total mass contained within a scale $L$ depends, in our approach, on the index $x(L)$ due to the self-similarity assumption (cf. equations \ref{eq_n-l} and \ref{eq_dens_scaling}): $M(L)=\langle \rho\rangle L^3 \propto L^{3/(1-x)}$. Hence, the mass-scale power-law index $\gamma\equiv3/(1-x)$ or: 
\begin{equation}
\label{gamma_x_local}
 x=\frac{\gamma-3}{\gamma}~.
\end{equation}

\begin{figure*} 
\begin{center}
\includegraphics[width=0.75\textwidth]{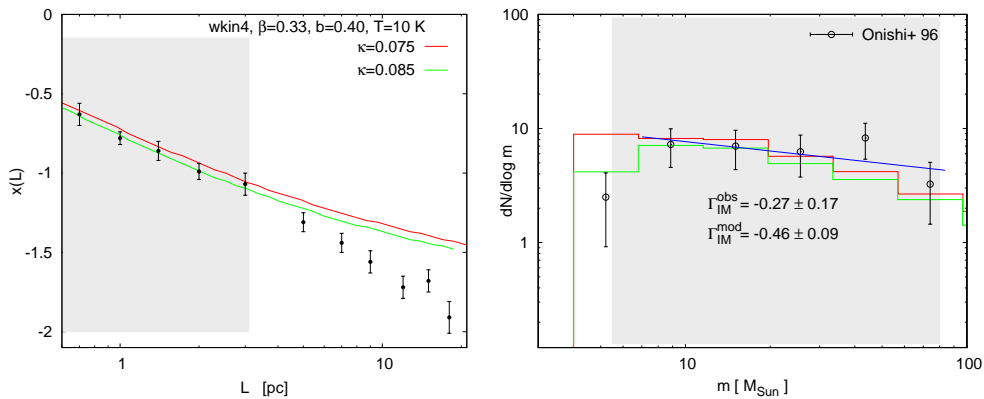}
\vspace{0.8cm}  
\caption{Structure and ClMF of the Taurus complex. {\it Left:} Best fits $x(L)$ of the data from LAL10 (dots with error bars) are plotted with lines and their characteristics (equipartition relation, model parameter values) are specified. {\it Right:} Clump mass functions, derived from the same models (lines), are juxtaposed with the observational ClMF from molecular line maps (open symbols with error bars). The range of scales of clump generation (left) corresponding to the observational clump mass range, restricted by the lower limit of confidence of the model (right) are shown with shaded areas. A weighted power-law fit of the observational ClMF (blue line) is drawn. Its slope (obs) and the one of the modeled intermediate-mass ClMF (mod) are specified. }
\label{fig_Taurus}
\end{center}
\end{figure*}

Comparison between $x(2R_{\rm s})$ obtained in that way from the work of LAL10 and our model predictions $x(L)$ for sets ($\beta$, $b$, $\kappa$, $T$) is described in Paper I (see Fig. 1 there). The best fits $x(L)$ for the cloud complexes Taurus, Orion A and Orion B studied in this Paper are plotted in the left panels of Fig. \ref{fig_Taurus}, \ref{fig_OriA} and \ref{fig_OriB}, respectively. We comment them in Section \ref{Results}.

\subsection{Derivation of the ClMF}
\label{Derivation_ClMF}
A parameter set ($\beta$, $b$, $\kappa$, $T$) corresponding to the obtained best fit $x(L)$ of the LAL10 data for a chosen MC yields: i) a lognormal clump mass distribution $p_L(m)$; and ii) a measure of the total number of clumps $N_{\rm tot}(L)$ at each scale (DVK11, Eq. 17):
\begin{equation} 
\label{eq_N_clumps_formula}
 N_{\rm tot}(L)=\frac{1}{\kappa^3}\exp\Big(\sigma^2\times\frac{(x-1)(1-2x)}{2x^2}\Big)~. 
\end{equation}

The ClMF of an individual cloud is derived as a superposition of the lognormal clump mass distributions. Each of them is discretized through sets of weights $\{N_L(m_j)=p_L(m_j)N_{\rm tot}(L),~j=1,...,n \}$ where the mass range is centred at the peak of the distribution and the mass limits are determined from the requirement:
\begin{equation}
 \label{eq_N_clumps_sum}
 N_{\rm tot}(L)=\sum\limits_{j=1}^{n} N_L(m_j)~.
\end{equation}

Also, to construct the ClMF correctly, one must require mass conservation throughout the MC fractal structure \citep{Elme_97}. The whole cloud with size equal to the upper limit of the inertial range $L_{\rm max}=20$~pc is to contain ${\cal N}_L$ substructures of mass $M(L)$ at given scale $L$. As shown in Paper I (Sect. 3.2): 
\begin{equation}
\label{eq_norm_Ntot}
 {\cal N}_L=\frac{L_{\rm max}^\frac{3}{1-x(L_{\rm max})}} {L^\frac{3}{1-x(L)}}~.
\end{equation}

Eventually, one obtains for the ClMF value in a selected mass bin $m^\prime - \Delta m^\prime\le m\le m^\prime + \Delta m^\prime$:
\begin{equation}
\label{eq_ClMF}
 F_{\rm ClMF} (m^\prime) = \sum\limits_L {\cal N}_L \sum_{m} N_L(m)~.
\end{equation}

If only gravitationally unstable clumps are considered, one should take also into account their contraction in timescales given by the free-fall time $\tau_{\rm ff}\propto \rho^{-1/2}$. Their {\it time-weighted} mass function can be derived by permitting constant replenishment of the clump population and introducing a weighting factor of each scale of clump generation: $w(L,x(L))\propto\tau_{\rm ff}^{-1}\propto \langle \rho \rangle^{1/2}\propto L^{3x/2(1-x)}$ (cf. equations \ref{eq_Larson_2} and \ref{eq_dens_scaling}). Then the time-weighted ClMF is obtained as the number of substructures ${\cal N}_L$ in equation \ref{eq_ClMF} is modified by factor $w(L,x(L))/w(L_{\rm max},x(L_{\rm max}))$.

\section{Results} \label{Results}
We put to test the ClMF, derived in Paper I, using molecular line observations of nearby Galactic MC complexes and dust continuum studies of further 3 regions at kpc-scale distances, mapped also in CO lines (Table~\ref{table_mc}). Subsamples of the original clump data are considered taking into account the lower mass limit of confidence in our models and/or the observational completeness limit; the corresponding size and mass ranges are specified in columns 3 and 4.

\begin{table*}
\caption{Observational clump samples used for comparison with the predictions of our model. Regions that are studied {\it both} by use of molecular line emissions and dust continuum data are put in bold. The slope estimates are re-derived by us, adopting mass limits corresponding to the applicability of our model and to the data completeness range. Notation: D = approximate distance to the region, IM = Intermediate-mass, HM = High-mass, $\gamma^{\rm obs}$ = Slope [$d\log m/d\log r$] on the clump mass-size diagram, $M_{\rm ch}$ = characteristic mass, CF = Clump-finding (method), {\sc pHM} = contour at Half-Maximum around intensity peaks, {\sc pPV} = Position-Velocity diagrams passing through intensity peaks; GB = gravitationally bound, UB = unbound}
\label{table_mc} 
\begin{center}
\begin{tabular}{lcccccccccc}
\hline 
\hline 
SF region & D & Ref. & Sizes & Masses & $\gamma_{\rm obs}$ & $M_{\rm ch}$& \multicolumn{2}{c} {ClMF slope} & CF method & Note   \\ 
~ & [ kpc ]  & ~ & [ pc ]  & [\Msol] & ~ & [\Msol] & IM & HM &  ~ & ~\\ 
\hline 
\multicolumn{11}{l} {\it Molecular line studies} \vspace{4pt}\\
Taurus & 0.14 & 1 & $0.07 - 0.45$ & $3 - 80$ &  $1.95$ & $\gtrsim100$ & $-0.27 $ & -- & {\sc pHM} & mostly GB \\
Orion A & 0.45 & 2 & $0.06 - 0.35$ & $30 - 1000$ &  $1.14$ & $\sim300$ & $-0.14$ & $-1.82$ & {\sc pPV} & $\sim$90\% starless, GB?  \\
Orion B & 0.45 & 3 & $0.06 - 0.70$ & $3 - 200$ &  $1.95$ & $\sim13$ & $-0.17$ & $-1.28$ & {\sc Gaussclumps} & UB + GB \\
MC sample & $\le0.18$ & 4 & $0.08 - 0.45$ & $3 - 60$ &  $2.09$ & $\sim10$ & $-0.25$ & $-1.45$ & various & mainly starless, GB? \\ 
~ & $\le0.18$ & 4 & $0.08 - 0.45$ & $3 - 100$ &  $2.09$ & $\sim13$ & $-0.44$ & $-1.77$ & various & mainly starless, GB? \\ 
{\bf M\,17} & 1.6 & 5 & $0.03 - 0.60$ & $6 - 2300$ &  $1.77$ & $\gtrsim200$ & $-0.34$ & $-1.00$ & {\sc Gaussclumps} & UB + GB? \\ 
{\bf Rosette} & 1.6 & 6 & $0.45 - 4.5$ & $30 - 2000$ &  $2.40$ & $\gtrsim800$ & $-0.28$ & $-1.13$ & {\sc Clumpfind} & starless, mostly UB? \vspace{6pt}\\ 
\multicolumn{11}{l} {\it Dust continuum studies} \vspace{2pt}\\
{\bf M\,17} & 1.6 & 7 & $0.02 - 0.70$ & $1 - 600$ &  $2.13$ & $\sim100$? & $-0.35$ & -- & {\sc Clumpfind} & mainly starless \\ 
\multicolumn{2}{l}{{\bf M\,17} + NGC\,7538} & 7, 8 & $0.06 - 0.60$ & $3 - 1000$ &  $2.05$ & $\sim100$ & $-0.25$ & $-1.82$ & {\sc Clumpfind} & UB + GB? \\ 
{\bf Rosette} & 1.6 & 9 & $0.20 - 0.70$ & $1.7 - 200$ & $2.22$ & $\gtrsim9$ & $-0.57$ & $-1.04$ & {\sc Getsources} & starless \\ 
\hline 
\hline 
\end{tabular} 
\end{center}
[1] \citet{Oni_ea_96}; [2] \citet{Tatematsu_ea_93}; [3] \citet{Kramer_ea_98}; [4] \citet{Tachi_ea_02}; [5] \citet{SG_90}; [6] \citet{WBS_95}; [7] \citet{RW_06a}; [8] \citet{RW_05}; [9] \citet{DiFran_ea_10}~~~~~~~~~~~~~~~~~~~~~~~~~~~\,\\
\smallskip 
\end{table*}

\begin{figure*} 
\begin{center}
\includegraphics[width=0.75\textwidth]{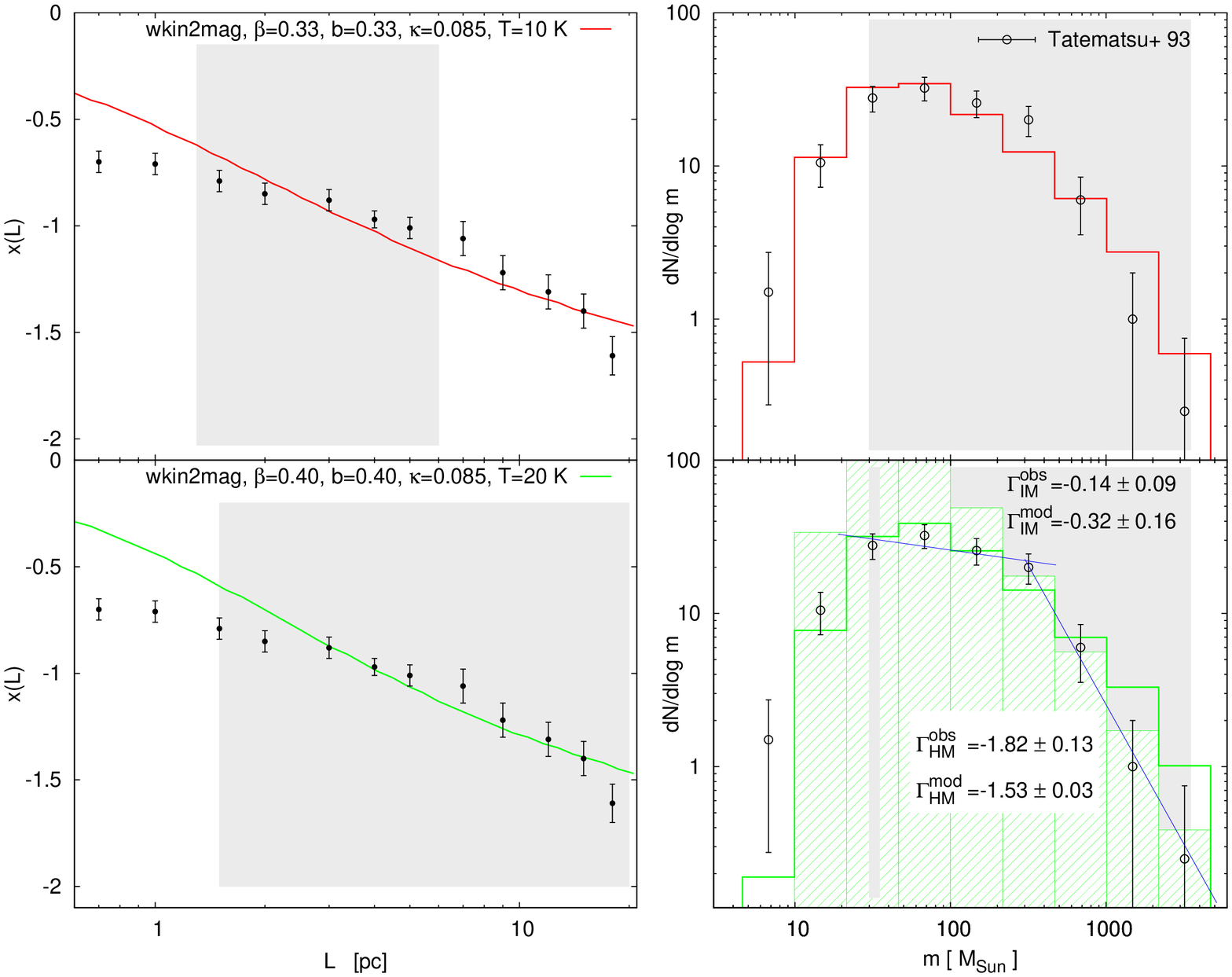}
\vspace{0.8cm}  
\caption{Structure and ClMF of the Orion A complex. The designations are the same like in Fig. \ref{fig_Taurus}; slopes of both intermediate-mass and high-mass ClMF are specified. The modeled non-time-weighted (lines) and time-weighted ClMFs (hatched area) are shown. The completeness limit of the observational data is adopted as a lower mass limit of confidence.}
\label{fig_OriA}
\end{center}
\end{figure*}

\subsection{Molecular line studies}

\begin{figure*} 
\begin{center}
\includegraphics[width=0.75\textwidth]{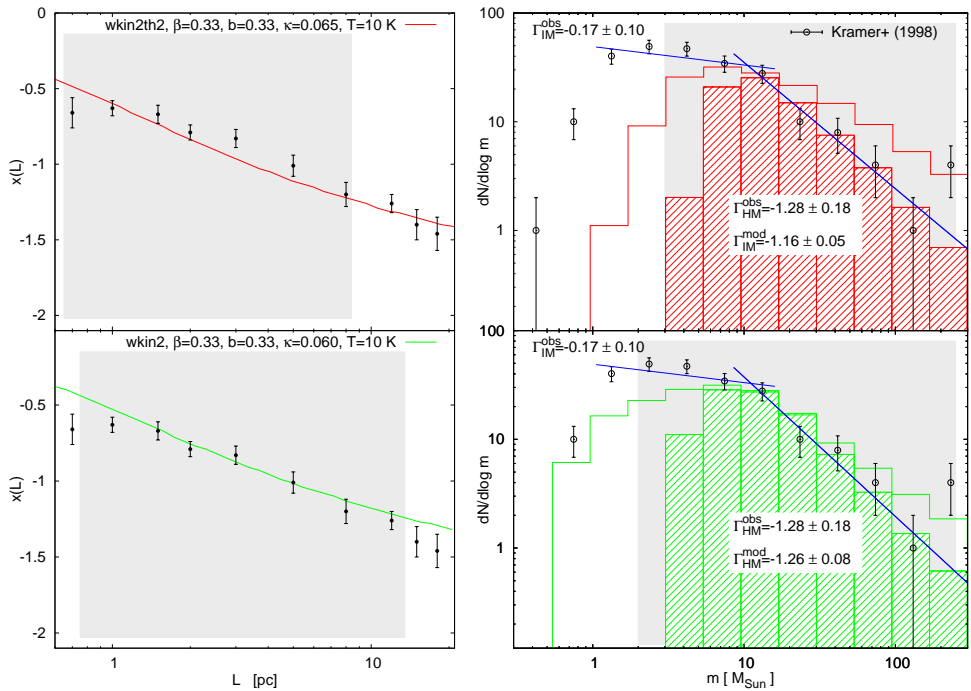}
\vspace{0.8cm}  
\caption{Structure and ClMF of the Orion B complex. The designations are the same like in Fig. \ref{fig_OriA}.}
\label{fig_OriB}
\end{center}
\end{figure*}

Three individual nearby star-forming regions of comparable size are selected: i) with clumps delineated through similar density tracers: \thirthCO \citep[Orion B;][]{Kramer_ea_98}, \CeightO \citep[Taurus;][]{Oni_ea_96} and \CS \citep[Orion A;][]{Tatematsu_ea_93}, consistent with the typical clump densities in our approach $n\sim 10^3-10^4$\cc; ii) with known general structure from dust extinction maps of LAL10. The spatial extent of the studied regions in the chosen complexes falls within the adopted inertial range of turbulence (Sect.~\ref{Physical_framework}, (i)). Dust extinction data reveal some diversity of structure in terms of mass-size relation of regions delineated by isodensity contours: Taurus exhibits monotonically `steep' structure in terms of $x(L)$ while Orion A and Orion B have `shallow' internal regions and `steep' external, less dense regions (Fig. \ref{fig_Taurus}-\ref{fig_OriB}, left panels; cf. Fig. 2 in LAL10). Taurus is a typical low-mass star-forming region \citep{KGW_08}, 
while numerous high-mass stars have been formed in Orion A and B in the recent past. As an additional test of the predictive power of the models for a variety of environments, we perform comparisons with observational ClMF for a sample of 9 MC complexes \citep{Tachi_ea_02} wherein star-forming, cluster-forming and starless clumps have been detected.

An equipartition relation and a parameter set were sought that yield the best fit $x(L)$ of the cloud structure as traced by the LAL10 data (Fig. \ref{fig_Taurus}-\ref{fig_OriB}, left panels). The modeled ClMF derived from them is compared with the observational one from molecular line maps (Fig. \ref{fig_Taurus}-\ref{fig_OriB}, right panels). The range of spatial scales considered in the fitting procedure (shaded areas, left panels) corresponds to the mass range of the referred observational ClMF (shaded areas, right panels), additionally restricted by the lower clump mass limit of confidence in the model. The scales are treated as those where the observed clumps were generated and hence their range is obtained through the clump mass-scale diagram of the tested model (Paper I, Sect. 3.1).

\subsubsection{Taurus}
\label{sect_Taurus}
Fig.~\ref{fig_Taurus} (left) demonstrates that the equipartition choice \texttt{wkin4} yields a very good fit of the inner parts of the complex. That seems physically consistent to us since this case describes structures which have been shaped under significantly influence of self-gravity. The typical dense clumps generated in them have mass distributions in a relatively good agreement with the observational ClMF (Fig.~\ref{fig_Taurus}, right), obtained by \citet{Oni_ea_96}. The best-fit parameters point to a mainly solenoidal turbulent forcing \citep[see][]{FKS_08} and a `soft' velocity scaling, typical for incompressible turbulence. On the other hand, the equipartition relation \texttt{wkin4} is obviously not applicable to the less dense extensive parts of the cloud. We stress, however, that in our model scales $L>3$~pc do not produce clumps with masses within the considered observational range.

It should be pointed out that the ClMF obtained by \citet{Oni_ea_96} differs significantly from ClMFs in other clouds as derived from molecular line emission data. Using essentially the lowest mass bin ($\simeq3$\Msol), these authors argue for a flat mass function\footnote{~A shallow mass spectrum, in their terms.} in Taurus, although they remain open for an alternative interpretation based on an ``unbiased survey... in a wide mass range''. We believe that indeed their data cover the mass range around the turn-over of the ClMF while its high-mass part (if existing) falls beyond their scope -- note that characteristic masses $M_{\rm ch}$ of hundreds\Msol~are derived in various MC complexes (Table 1, column 7). The power-law slope of the modeled ClMF over the lower mass limit of confidence in our model is slightly shallower than the typical one of the mass distribution of CO clumps \citep{Blitz_93, Kramer_ea_98}. It would steepen ($\Gamma\sim-0.7$) if the observational data contained clumps with masses up 
to the expected $M_{\rm ch}\sim200$\Msol~  -- cf. the modeled ClMFs from \wkiniv~(Fig. 3 in Paper I).

\subsubsection{Orion A}
\label{sect_Orion_A}
We obtained good fits of the Orion A structure in the case \texttt{wkin2mag} at intermediate to large spatial scales (Fig. \ref{fig_OriA}, left). The corresponding ClMFs generally agree with the result of \citet{Tatematsu_ea_93}. Enhancement of the model temperature, compatible with the derived $T\sim20$~K in the complex \citep{ISK_07}, leads to larger, more plausible values of $b$ and $\beta$ and yields a wider range of scales of clump generation as the cloud structure is better fitted through the curve $x(L)$ (Fig. \ref{fig_OriA}, bottom).

Depending on the model temperature, the case \texttt{wkin2mag} in our approach generates mostly/only gravitationally unstable clumps (Paper I, Fig. 3). On the other hand, \citet{Tatematsu_ea_93} claim that all clumps in their sample are close to virial equilibrium. Therefore a time-weighted ClMF (cf. Sect.~\ref{Derivation_ClMF}) is more appropriate for comparison with the referred observational study. Time-weighting steepens the high-mass slope of the model from $\Gamma{\rm^{mod}_{HM}}\sim-1$, typical for fractal clouds, to $\sim-1.5$, i.e. steeper than that of the stellar IMF \citep[$\Gamma\simeq-1.3$;][]{Sal55} (Fig.~\ref{fig_OriA}, bottom right). This leads to agreement (within the data uncertainties) with the observational ClMF for clump masses $>M_{\rm ch}\sim300$\Msol~ while for lower masses the deviation is drastic. In contrast, the non-weighted modeled ClMF is generally consistent with the observational ClMF below $M_{\rm ch}$ and for $T=20$~K. We note the high completeness limit ($\sim30$\Msol)
~of the data which might artificially constrain the intermediate-mass range.

\subsubsection{Orion B}
\label{sect_Orion_B}
The referred observational ClMF in this region is part of the work of \citet{Kramer_ea_98} who analyzed CO data sets for 8 Galactic MCs. They argued for a single power-law ClMFs in all cases, with a universal mean slope $\Gamma\sim-0.7$, over a wide range of clump masses. However, a closer look at their result for Orion B reveals that this MC may be an exception. The conclusion of \citet{Kramer_ea_98} about a single power-law with shallow slope is based mostly on the single mass bin $\gtrsim200$\Msol~(Fig.~\ref{fig_OriB}, right). Inspection of the statistically rich data in the range $2-80$\Msol~indicate rather a combination of two power laws: a shallow intermediate-mass slope below $15$\Msol~and a significantly steeper high-mass slope. Adopting the latter value as $M_{\rm ch}$ and the completeness data limit of the authors $M\sim1$\Msol, one obtains $\Gamma{\rm_{IM}^{obs}}$ similar to that in Orion A and a Salpeter-like $\Gamma{\rm_{HM}^{obs}}$.

\begin{figure}
\begin{center}
\includegraphics[width=84mm]{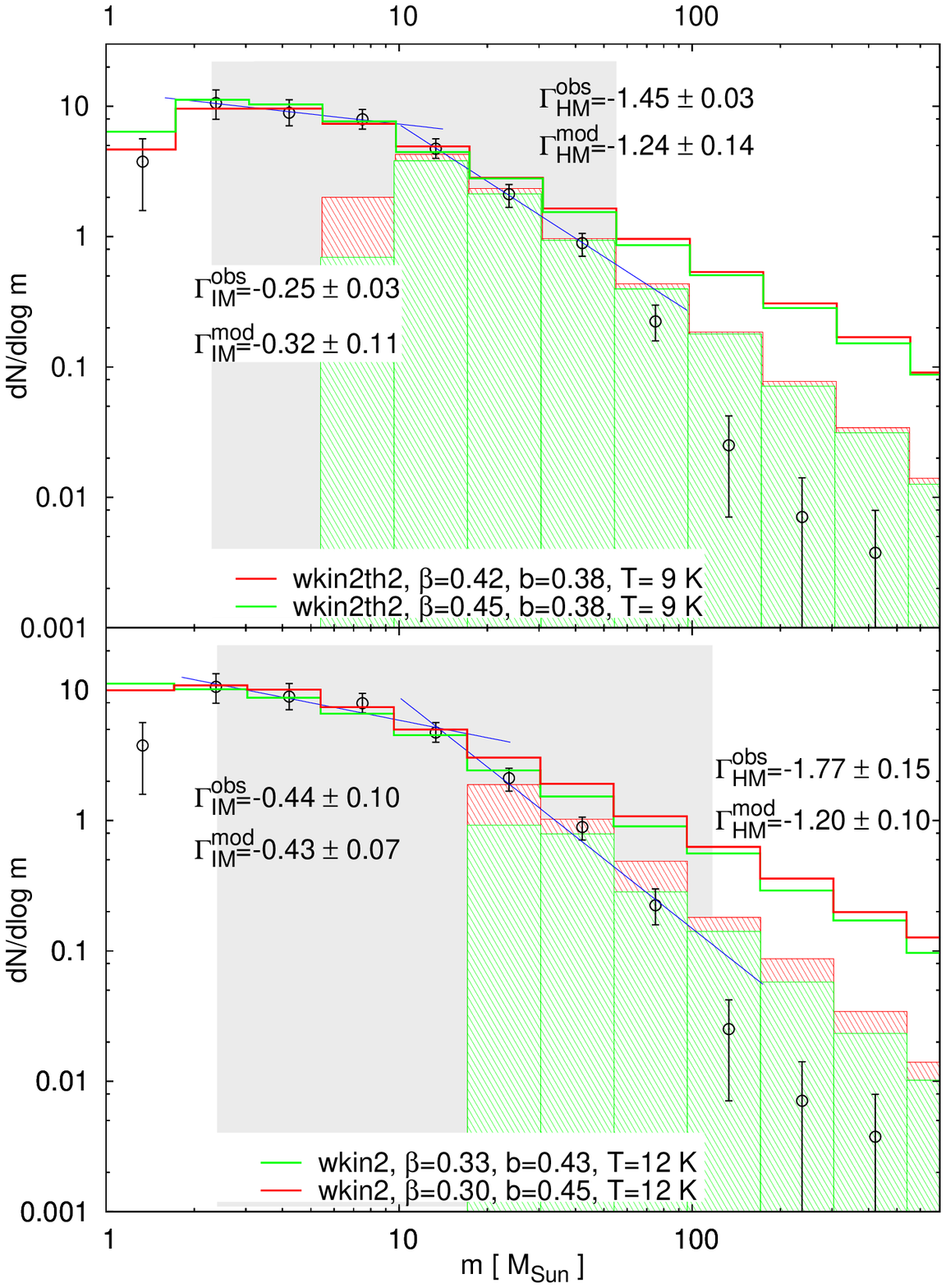}
\vspace{0.7cm}  
\caption{Modeling of the observational ClMF from 9 Galactic star-forming regions by use of the equipartition relations \wkinii~and \wkinth~for $\kappa=0.065$ and adopting characteristic mass $\sim10$\Msol~(top) and $\sim13$\Msol~(bottom). See text for the considered mass ranges. Other designations are the same like in Figs.~\ref{fig_Taurus}-\ref{fig_OriB}.}
\label{fig_MCs}
\end{center}
\end{figure}

Modeling of the Orion B structure as traced by the LAL10 data yields good fits in the cases \texttt{wkin2th2} and \texttt{wkin2} (Fig.~\ref{fig_OriB}, left). For such choices of equipartition and temperature $T=10$~K, the critical mass over which most of the generated clumps are gravitational unstable is about the adopted characteristic mass from the used observational data. Therefore we apply time-weighting for the high-mass regime and obtain $\Gamma{\rm_{HM}^{mod}}$ close to the Salpeter value. The modeled ClMFs fit remarkably well the high-mass observational ones, excluding the sole bin at the upper limit of the distribution (Fig.~\ref{fig_OriB}, right). Non-weighted ClMFs obviously fail to reproduce the observational one although the high-mass slope $\sim-0.7$ is close to the estimate of \citet{Kramer_ea_98}. The model is not applicable to fit the intermediate-mass ClMF because of the high lower mass limit of confidence.

\subsubsection{Sample of MC complexes}
The presented best-fit descriptions $x(L)$ of MC structure in Taurus, Orion A and Orion B and their corresponding ClMFs with different characteristic masses $M_{\rm ch}$ and HM slopes (cf. Table~\ref{table_mc}) are derived by use of different equipartition relations that may reflect a variety of physical conditions. What model case would describe appropriately a statistical ClMF obtained from a large set of clump data in different MCs? The sample in the study of \citet{Tachi_ea_02} could give a clue to the answer -- it encompasses nearby cloud complexes of diverse star-forming type as the vast majority of clumps are associated with no or a small number young stellar objects and have masses below several dozens\Msol. Clumps associated with young stellar clusters were found in two out of nine sampled clouds and their fraction is only 5 \% of the total number of clumps. All objects with masses $\ge115$\Msol~are from this group. One may take the latter value as an upper mass limit when fitting 
the ClMF, in order to avoid extreme environments and to secure sufficient statistics. The two power-law shape of the ClMF is evident whereas the characteristic mass is ambiguous. We consider two choices which both allow for very good fits: i) $M_{\rm ch}\sim 10$\Msol, adopting the estimate of \citet{Tachi_ea_02} and decreasing the upper mass limit further by one bin; and ii) $M_{\rm ch}\sim 13$\Msol (Fig.~\ref{fig_MCs}). The lower mass limit in both cases is the lower limit of confidence of the tested models.

The choice i) yields excellent agreement of the model case \wkinth~with the observational ClMF: time-weighted model for the HM part and non-time-weighted for the IM part (Fig.~\ref{fig_MCs}, top). The IM slope is very similar to the ones found in the considered individual MC complexes while the HM slope is about the Salpeter value, like in Orion B (Sect.~\ref{sect_Orion_B}). The discrepancy with the data for $m\gtrsim60$\Msol~could be attributed to incompleteness of the sample in this mass range. A Salpeter-like slope is to be expected for mass distributions of (nearly) virialized clumps assuming their formation at a constant rate. The effect will be similar if the ClMF is derived from a statistically significant sample containing clumps in complexes at different evolutionary stages of clump formation. Note also that the variation of the velocity scaling index $\beta$ in the fitting models is consistent with most observational and numerical works \citep[e.g.][]{Pad_ea_06, Pad_ea_09}. The values of the 
$b$ parameter indicate preliminary solenoidal turbulent forcing with small contributions of the compressive mode.   

Considering the choice ii), an upper mass limit $\sim115$\Msol~is adopted as mentioned above. The model case \texttt{wkin2} provides an excellent fit of the intermediate-mass ClMF and a problematic one -- for the high-mass ClMF. Again, the completeness of the clump mass bin $60\le m\le100$\Msol~is an open issue which is crucial for the correct slope estimation. On the other hand, the obtained low velocity scaling index $\beta$ is consistent with results for Taurus, Orion A and Orion B.  

Modeling of the observational ClMF from the equipartition cases \wkinii~and \wkinth~suggests that most of the clumps generated in MC complexes of diverse star-forming activity and with mean density about the typical local density (cf. Eq. 7 in Paper I) are in a state close to virial equilibrium.  

\begin{figure*} 
\begin{center}
\includegraphics[width=0.75\textwidth]{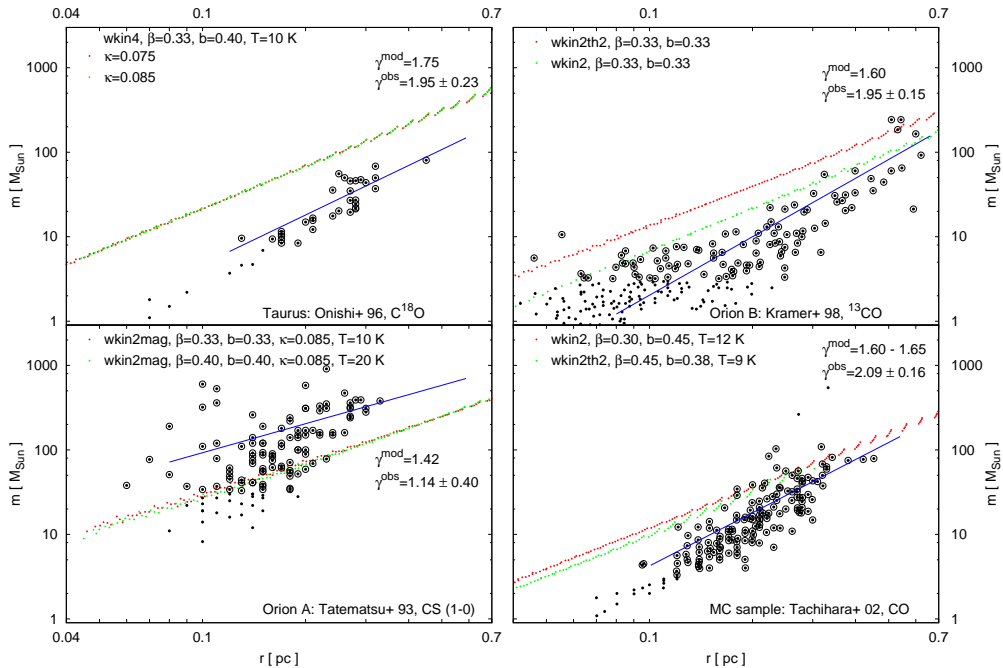}
\vspace{0.7cm}  
\caption{Clump mass-size diagrams from the referred molecular line studies. The data used to derive the ClMFs are shown with open symbols. The predictions of our model are plotted with dots, retaining the corresponding colors from Figs. \ref{fig_Taurus}-\ref{fig_MCs}.}
\label{fig_ML_MR}
\end{center}
\end{figure*}

\subsubsection{Clump mass-size diagrams}
\label{mr_diagrams}
An additional test for adequacy of the proposed statistical modeling of the ClMF is to construct clump mass-size diagrams. The comparison between the used observational data and the best-fit models of the observational ClMFs is displayed in Fig. \ref{fig_ML_MR}. There are two difficulties with such analysis. First, the `average clump ensembles' in our model occupy narrow strips on $m-r$ diagrams\footnote{~Clump sizes from our model are converted to radii $r=l/2$ to be compared with the radii of observed clumps.} whereas the dispersion of the observational samples is huge. Second, mass and (especially) size determinations from molecular-line maps depend essentially on the clump-finding technique which is different in each considered reference work (Table~\ref{table_mc}, column 10). A meaningful criterion is to compare the slope $\gamma^{\rm obs}=(d\log m/d\log r)$ on the observational $m-r$ diagram with the predicted slopes $\gamma^{\rm mod}$ as the latter vary with the chosen equipartition case and the 
velocity scaling index $\beta$ (cf. Paper I, Fig. C1).

The predicted slopes from the best-fit ClMF model are consistent with the ones from the referred molecular-line studies within $1\sigma$ limit for Taurus and Orion A and within about $2\sigma$ limit for Orion B and for the sample of \citet{Tachi_ea_02}. In two cases, the `average clump ensemble' strips cross the region of the observational data (Fig.~\ref{fig_ML_MR}, bottom). Taking into account the good agreement between observational and modeled ClMFs (Figs. \ref{fig_Taurus}-\ref{fig_MCs}), the shift between the two sets on the $m-r$ diagrams -- with a factor of 2 toward larger (Orion A) or smaller sizes, -- is to be interpreted with the variety of the clump {\it size} definition. The issue will be commented further in the next Section wherein we compare our model with ClMFs from dust continuum studies and in the Discussion.

\subsection{Dust continuum studies}
\label{sect_dust_continuum}
Sensitive submillimeter continuum maps are appropriate for probing the MC structure on small scales due to the optically thin dust emission. In the last decade, the achieved angular resolutions of $\lesssim20^{\prime\prime}$ from ground-based observations (e.g. SCUBA, Bolocam) and space missions like {\sc Herschel} allowed for study of clumps with sizes $\gtrsim 0.1$~pc in complexes at distances over $1$~kpc. Therefore it is instructive to compare results from such studies with our model. In view of the complexes' remoteness, the considered regions were not included in the work of LAL10 and we don't have at our disposal their dust extinction mapping as a tool to trace and fit the general cloud structure (cf. Figs. \ref{fig_Taurus}-\ref{fig_OriB}, left). To compensate this lack, we chose star-forming regions which clump population has been probed also by use of molecular-line maps. The observational ClMF was fitted directly and the clump mass-size diagram was used as an additional test.

\begin{figure*} 
\begin{center}
\includegraphics[width=0.75\textwidth]{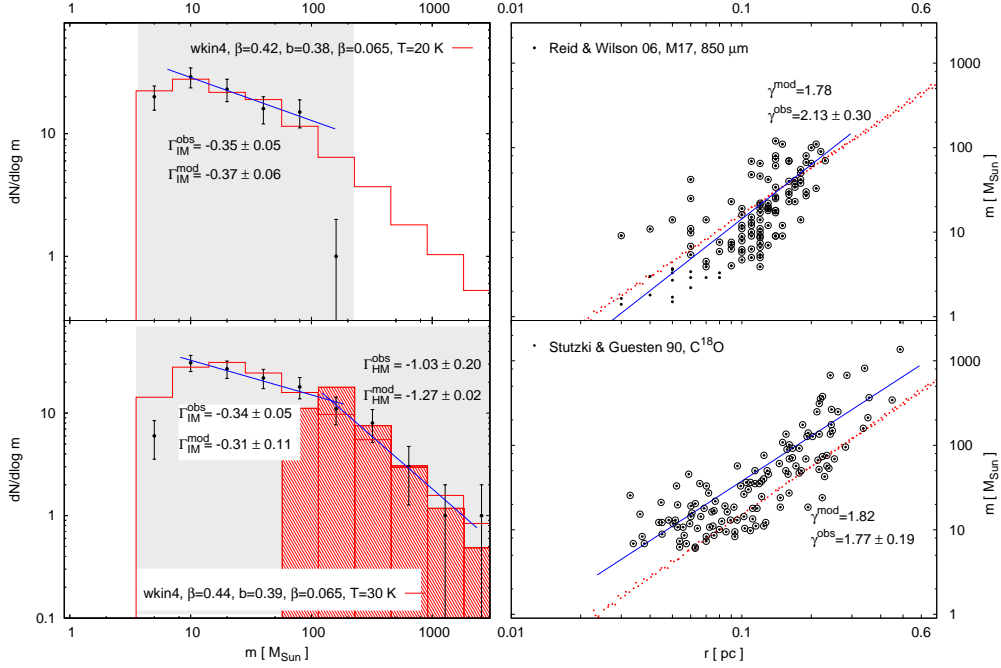}
\vspace{0.7cm}  
\caption{ClMF and clump mass-size diagrams of the star-forming region M\,17 from dust-continuum (top) and molecular-line (bottom) observations. The shaded areas denote the observational mass range restrained by the lower-mass confidence limit from the model -- the corresponding data used to derive the ClMFs are shown with open symbols in the M-R diagram. The predictions of the model are plotted with the same colors in the left (lines) and right (dots) panels.}
\label{fig_M17}
\end{center}
\end{figure*}

\subsubsection{M\,17}  
Large part of this massive star-forming region was mapped by \citet{RW_06a} in two SCUBA bands. We use here the clump data from their $850~\mu$m map because of the richer statistics. The fitting of the observational ClMF and the corresponding clump-mass size diagram are plotted in Fig. \ref{fig_M17} (top). A very good fit of the ClMF is achieved from the model case \wkiniv~that points to strongly gravitationally bound clumps. The predicted and the observational clump mass-size diagrams are also in agreement (Fig. \ref{fig_M17}, right top). Note that we assumed a higher gas temperature $T=20$~K which is about the average clump temperature derived by the authors. 

Although \citet{RW_06a} suggest a two power-law ClMF with $M_{\rm ch}\sim8$\Msol, it seems that their result hints rather at an order-of-magnitude higher characteristic mass. This is supported by the slope in the mass range $10\lesssim m \lesssim100$\Msol~which is similar to the IM slopes found from the referred molecular-line studies (cf. Table~\ref{table_mc}). However, only two clumps with masses $>100$\Msol~in the observational sample are far from sufficient for a plausible estimate of the characteristic mass. To clarify the issue, we apply our method to a \CeightO study of the south-western sector of M\,17 \citep{SG_90}. As demonstrated in Fig.~\ref{fig_M17}, bottom, the results for an intermediate-mass ClMF with $M_{\rm ch}\sim100$\Msol~are virtually the same, fitted from the same model case and free parameter values and slightly higher temperature. Moreover, the high-mass ClMF is also well fitted by the time-weighted mass distribution of gravitationally unstable clumps. Neglecting the mass bin $>2000$\
Msol, which contains only 2 clumps, the slope $\Gamma_{\rm HM}$ is shallower than that of the stellar IMF.   

\begin{figure*} 
\begin{center}
\includegraphics[width=0.75\textwidth]{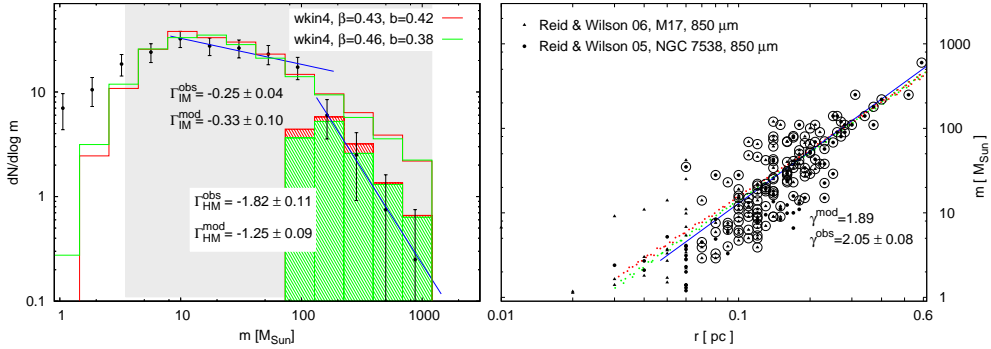}
\vspace{0.7cm}  
\caption{Combined ClMF and clump mass-size diagram of the star-forming regions M\,17 and NGC\,7538 from dust-continuum observations. The designations are the same like in Fig. \ref{fig_M17}. The values of $\kappa=0.065$ and $T=30$~K are retained without change from the fitting of the M\,17 sample only.}
\label{fig_M17_NGC7538}
\end{center}
\end{figure*}

For independent confirmation of this result, we composed a twice larger sample, including clump data from a similar study of NGC\,7538 \citep{RW_05}. As shown in Fig. \ref{fig_M17_NGC7538}, the ClMF is fitted again from the model case \wkiniv~and a similar parameter set $(\beta, b)$ for a large range of clump masses $7\lesssim m \lesssim1000$\Msol, applying time-weighting for the more massive clumps. In comparison to the mass distribution from the M\,17 sample, the IM slope shallows slightly and approaches the values from the molecular-line studies of nearby MC complexes (cf. Table~\ref{table_mc}) while the HM slope is much steeper than that from Stuzki \& G\"usten's study and is identical with the one in Orion A (Fig.~\ref{fig_OriA}) although with a smaller $M_{\rm ch}$. The agreement of the model with the clump data on the $m-r$ diagram is again excellent (Fig. \ref{fig_M17_NGC7538}, right).

\subsubsection{Rosette molecular cloud}
\begin{figure*} 
\begin{center}
\includegraphics[width=0.75\textwidth]{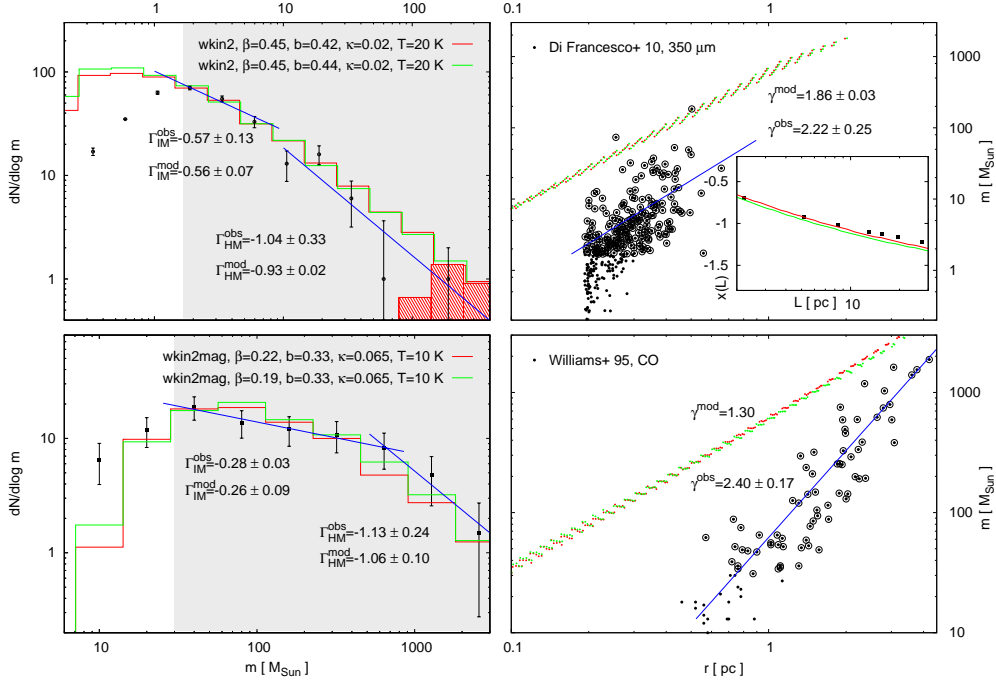}
\vspace{0.7cm}  
\caption{ClMF and clump mass-size diagrams of the Rosette MC from dust-continuum (top) and molecular-line (bottom) observations. The designations are the same like in Fig. \ref{fig_M17}. A comparison between the predictions of the illustrated models and the observed structure from a column density map using Herschel FIR-continuum data (N. Schneider, priv. comm.) is embedded in the top-right panel.}
\label{fig_Rosette}
\end{center}
\end{figure*}

The abundant observational studies of this complex in the outer Galaxy indicated active star formation in the past and nowadays. Its clump population has been studied both on molecular-line and dust continuum maps; we chose the works of \citet{WBS_95} and \citet{DiFran_ea_10}, respectively. The high-quality {\sc Herschel} data used by the latter authors allow for derivation of the ClMF in a mass range spanning three orders of magnitude (Fig. \ref{fig_Rosette}, top left). Some incompleteness is sensible in the mass bins $\gtrsim10$\Msol~and has -- in our view, -- two possible explanations. First, for the sake of our study we selected only starless clumps from the original sample. Second, \citet{DiFran_ea_10} applied the {\sc Getsources} algorithm \citep{Mensh_ea_12} for clump decomposition which favors identification of more compact (and less massive) objects. 

Because of the lower mass-limit of confidence, our statistical method is not able to fit the ClMF below $\sim2$\Msol. The observed mass distribution for larger masses could be interpreted as single power-law ClMF with a slope about $-1$ (not shown). In view of the possible incompleteness of the high-mass clump data, we suggest rather a two power-law mass function, fitted from the model case \wkinii~(Fig. \ref{fig_Rosette}, top left). The characteristic mass for \wkinii~and for the obtained best-fit parameters is $\sim9$\Msol~(see Fig. 3 in Paper I). Adopting this value for $M_{\rm ch}$, one gets a bit steeper intermediate-mass ClMF than found for other MC complexes in this Paper and a high-mass ClMF, with a slope typical for fractal clouds \citep{Elme_97}. On the clump mass-size diagram, a shift of the model from the observed clumps is evident like in some other studied regions: by a factor of 2 to 4 (Fig. \ref{fig_Rosette}, top right; cf. Fig. \ref{fig_ML_MR}). Nicola Schneider (private communication) provided for us estimates of the cloud mass $M$ as a function of spatial scale $L$, calculated from the original {\sc Herschel} column density map of Rosette. That enabled a check of the derived best-fit model with the observed cloud structure, in terms of the LAL10 work (Figs. \ref{fig_Taurus}-\ref{fig_OriB}, left). The shape of the modeled curve $x(L)$ is consistent with the dust-continuum data but with a shift toward lower $x$. Decreasing the mapping resolution parameter down to $\kappa=0.02$, we found excellent agreement with the data (Fig. \ref{fig_Rosette}, right, embedded diagram). That is illustrative how the observed general cloud structure affects the $\kappa$-parameter space of the ClMF fitting. 

\citet{WBS_95} derived the observational ClMF from CO mapping of Rosette MC. As seen in Fig. \ref{fig_Rosette} (bottom), their clump size range practically does not overlap with the one of \citet{DiFran_ea_10} while the mass range spans from $10$ to $>1000$\Msol. The ClMF is well fitted from the model case \wkinmag~and a very `soft' velocity scaling, $\beta\simeq0.20$. Such values of the scaling index were found from magneto-hydrodynamic simulations of \citet{Collins_ea_12}. 

\section{Discussion}	\label{Discussion}
The comparison with the considered molecular-line and dust-continuum studies of Galactic MC complexes demonstrates the applicability of a statistical approach to derive the ClMF presented in Paper I. The results on fitting the observational ClMF through our model are summarized in Table \ref{table_fit}. They are obtained by use of two different primary fitting criteria: fitting of the general cloud structure $x(L)-L$ or direct fitting of the ClMF. Regardless of the type of observational data, the applied fitting criterion and of the variety of model cases, the best-fit parameters span relatively narrow ranges: 
\begin{itemize}
 \item `Soft' velocity scaling: $0.20\lesssim \beta \le 0.46$
 \item Mainly solenoidal forcing: $0.30 \le b \le 0.46$
 \item Mapping resolution from few percent to one tenth of the spatial scale: $0.02 \le \kappa \le 0.10$
 \item Typical temperatures for molecular gas phase: $8\le T\le 35$~K.
\end{itemize}
A small velocity scaling index $\beta$, close or equal to the Kolmogorov value for incompressible turbulence ($0.33$), may seem unrealistic in view of the high compressibility of interstellar turbulence which implies $\beta\sim0.50$. In fact, numerical simulations of magnetized clouds show that the velocity power spectrum can be even shallower than in the Kolmogorov theory. \citet{Collins_ea_12} measured $\beta=0.23-0.29$ for thermal-to-magnetic pressure ratio in range $0.2-2.0$. (In our modeling, the latter value could be as low as $\simeq0.05$ in case \texttt{wkin2mag}; see Eq.~\ref{eq_mag_scaling}.) Apparently, the velocity scaling index for incompressible turbulence $\beta=0.33$ should not be necessarily treated as the lowest possible value from theoretical considerations.

The best-fit range of the turbulent forcing parameter $b$ is narrower, with typical values $0.38-0.40$, in most considered complexes. Since $b$ can vary within a star-forming region, this result might be explained as a statistical effect. Indeed, $b\simeq0.40$ corresponds to a natural mixture between solenoidal and compressive modes as the latter represent longitudinal waves, occupying one of the three spatial dimensions \citep[see][and Fig. 8 there]{Fed_ea_10}. The typical best-fit values of the mapping resolution parameter $\kappa$ are about several percent. These are the expected values, appropriate to distinguish substructures which are significantly smaller than the spatial scale $L$ and significantly larger than the scale of dissipation.

Generally, our results lend support to a two power-law shape of the ClMF: intermediate-mass and high-mass part, with two distinct values of the characteristic mass $M_{\rm ch}$: $\sim10$ and $\sim200$\Msol. (The molecular-line study of Rosette MC with its high $M_{\rm ch}$ is the only exception.) The first value is consistent with model cases of a ``virial-like'' equipartition between gravitational and turbulent energy, possibly with contribution of the thermal energy at small scales and negligible contribution of the magnetic energy (\wkinii, \wkinth). These equipartition relations are not sufficient to argue that the modeled clumps are in virial equilibrium \citep{BP_06} but rather indicate their gravitational boundedness or contraction. On the other hand, if the considered model cases hold for a whole cloud, they are indicative for its global collapse \citep{VS_ea_07}. Large characteristic masses of hundreds \Msol~are obtained for strongly gravitating clumps, with possible magnetic support (\wkiniv, \wkinmag). Such are evidently the cases in Taurus, Orion A, M\,17 and NGC\,7538. 

The slope of the modeled intermediate-mass ClMF is shallow and does not vary significantly from complex to complex ($-0.26\ge \Gamma_{\rm IM}^{\rm mod}\ge-0.56$) while the variety is larger for the high-mass one: from typical slopes for fractal clouds to slopes, a bit steeper than the one of the stellar IMF ($-0.9\gtrsim \Gamma_{\rm HM}^{\rm mod}\gtrsim-1.5$). It seems that the single power-law observational ClMFs of slope $-0.7\gtrsim \Gamma^{\rm obs}\gtrsim -0.8$  derived from molecular-line studies in 1990s \citep{SG_90, Blitz_93, Heit_ea_98} have been products of a combination between intermediate- and a few bins from high-mass ClMF. On the other hand, the variety of high-mass ClMF slopes probably reflects a real variety of physical conditions in individual complexes and the gravitational balance and evolution of the dense fragments in them. Its model slope $\Gamma^{\rm mod}_{\rm HM}$ similar to that of the stellar IMF when the ClMF is time-weighted. When derived from molecular-line mapping of Orion A \citep{Tatematsu_ea_93} or the dust-continuum studies of M\,17 and NGC\,7538 by \citet{RW_05, RW_06a}, $\Gamma^{\rm obs}_{\rm HM}$ exceeds noticeably the Salpeter value but lies still within the observed variability range \citep{Kroupa_01}.

The dust extinction mapping of Taurus, Orion A and Orion B by LAL10 enabled us to make a link between the predicted general cloud structure and the ClMF. We point out also the excellent agreement applying this criterion to the dust-continuum study of Rosette, making use of a column density map of Rosette (Fig. \ref{fig_Rosette}, top right, embedded) obtained from {\sc Herschel data} \citep[see][for details]{Schneider_ea_12}. (Note that the extinction map derived from near-IR extinction using 2MASS \citep{Schneider_ea_11} delivers similar values for $A_{V}\lesssim10^m$.) Generally, good agreement is found for the chosen scales of clump generation $L$ between $\sim0.5$ and $\sim15$~pc while in Taurus the upper limit is about half of that. Those are the conservative limits of the inertial range of turbulence\footnote{~We recall here the comment in Sect. \ref{Physical_framework}, i)} that is a corner stone in our framework of clump description. The discrepancies for larger $L$ are to be expected and can be interpreted as reflecting changes in the physical conditions -- such scales may lie outside the inertial range and/or the assumption of isothermality does not hold for them \citep[e.g.][]{Henne_ea_08}. On the other hand, at $L\lesssim1$~pc the density PDF deviates from the lognormal shape and develops a power law tail in the high-density part \citep{Kless_00, KNW_11}. The predicted typical clump size in such density regimes approaches $\sim0.1$~pc which is about the sonic scale for temperature range $10-20$~K, i.e. the assumption for supersonic turbulence as a generator of clumps breaks down. 

\begin{table*}
\caption{Summary of the results on fitting the observational ClMF in the considered molecular cloud complexes. The probed parameter ranges (below the table) and the ranges of the model parameters that give good fits (Columns 3-7) are specified. Notation: FC = Fitting Criterion ($x-L$ -- general cloud structure on $x(L)-L$ diagram; ClMF -- direct fitting of the ClMF), MC = model case.}
\label{table_fit} 
\begin{center}
\begin{tabular}{@{\,}l@{~\,}c@{~\,}c@{~\,\,}c@{~\,\,}c@{~\,\,}c@{~\,}c@{~\,\,}c@{~\,}c@{~\,}c@{~\,}c@{~\,}c@{~}c}
\hline 
\hline 
SF region & FC & MC & \multicolumn{4}{c} {Model parameter ranges} & \multicolumn{4}{c} {ClMF} & \multicolumn{2}{c} {$m-r$ slope} \\
~ & ~  & ~ & $\beta$ & $b$ & $\kappa$ & $T$ [K] & $\Gamma^{\rm obs}_{\rm IM}$ & $\Gamma^{\rm mod}_{\rm IM}$ & $\Gamma^{\rm obs}_{\rm HM}$ & $\Gamma^{\rm mod}_{\rm HM}$ & $\gamma^{\rm obs}$ & $\gamma^{\rm mod}$ \\ 
\hline 
\multicolumn{13}{l} {\it Molecular line studies} \vspace{4pt}\\
Taurus & $x-L$ & \wkiniv & 0.30-0.36 & 0.38-0.42 & 0.07-0.09 & 10-20 & $-0.27{\scriptstyle \pm 0.17}$ & $-0.46{\scriptstyle \pm 0.09}$ & -- & -- & $1.95{\scriptstyle \pm 0.23}$ & $1.75$ \\
Orion A & $x-L$ & \wkinmag & 0.38-0.42 & 0.38-0.42 & 0.07-0.09 & 18-22 & $-0.14{\scriptstyle \pm 0.09}$ & $-0.32{\scriptstyle \pm 0.16}$ & $-1.82{\scriptstyle \pm 0.13}$ & $-1.53{\scriptstyle \pm 0.03}$ & $1.14{\scriptstyle \pm 0.40}$ & $1.42$ \\
Orion B & $x-L$ & \wkinii & 0.30-0.36 & 0.30-0.36 & 0.055-0.065 & 10-15 & $-0.17{\scriptstyle \pm 0.10}$ & -- & $-1.28{\scriptstyle \pm 0.18}$ & $-1.26{\scriptstyle \pm 0.08}$ & $1.95{\scriptstyle \pm 0.15}$ & $1.62$ \\
~ & $x-L$ & \wkinth & 0.30-0.36 & 0.30-0.36 & 0.050-0.065 & 10-15 & $-0.17{\scriptstyle \pm 0.10}$ & -- & $-1.28{\scriptstyle \pm 0.18}$ & $-1.16{\scriptstyle \pm 0.05}$ & $1.95{\scriptstyle \pm 0.15}$ & $1.60$ \\
MC sample & ClMF & \wkinii & 0.30-0.36 & 0.42-0.46 & 0.02-0.10 & 10-12 & $-0.44{\scriptstyle \pm 0.10}$ & $-0.43{\scriptstyle \pm 0.07}$ & $-1.77{\scriptstyle \pm 0.15}$ & $-1.20{\scriptstyle \pm 0.10}$ & $2.09{\scriptstyle \pm 0.16}$ & $1.60$ \\
~ & ClMF & \wkinth & 0.42-0.46 & 0.36-0.40 & 0.02-0.10 & 8-10 & $-0.25{\scriptstyle \pm 0.03}$ & $-0.32{\scriptstyle \pm 0.11}$ & $-1.45{\scriptstyle \pm 0.03}$ & $-1.24{\scriptstyle \pm 0.14}$ & $2.09{\scriptstyle \pm 0.16}$ & $1.65$ \\
{\bf M\,17} & ClMF & \wkiniv & 0.43-0.45 & 0.37-0.40 & 0.02-0.10 & 28-35 & $-0.34{\scriptstyle \pm 0.05}$ & $-0.31{\scriptstyle \pm 0.11}$ & $-1.03{\scriptstyle \pm 0.20}$ & $-1.27{\scriptstyle \pm 0.02}$ & $1.77{\scriptstyle \pm 0.19}$ & $1.82$ \\	
{\bf Rosette} & ClMF & \wkinmag & 0.18-0.23 & 0.30-0.36 & 0.02-0.10 & 8-15 & $-0.28{\scriptstyle \pm 0.03}$ & $-0.26{\scriptstyle \pm 0.09}$ & $-1.13{\scriptstyle \pm 0.24}$ & $-1.06{\scriptstyle \pm 0.10}$ & $2.40{\scriptstyle \pm 0.17}$ & $1.30$\vspace{6pt}\\	
\multicolumn{13}{l} {\it Dust continuum studies} \vspace{4pt}\\
{\bf M\,17} & ClMF & \wkiniv & 0.40-0.44 & 0.37-0.40 & 0.02-0.10 & 28-35 & $-0.35{\scriptstyle \pm 0.05}$ & $-0.37{	\scriptstyle \pm 0.06}$ & -- & -- & $2.13{\scriptstyle \pm 0.30}$ & $1.78$ \\	
{\bf M\,17}+ & ClMF & \wkiniv & 0.43-0.46 & 0.38-0.42 & 0.02-0.10 & 28-35 & $-0.25{\scriptstyle \pm 0.04}$ & $-0.33{	\scriptstyle \pm 0.10}$ & $-1.82{\scriptstyle \pm 0.11}$ & $-1.25{\scriptstyle \pm 0.09}$ & $2.05{\scriptstyle \pm 0.08}$ & $1.89$ \\	
{\bf Rosette} & ClMF & \wkinii & 0.44-0.46 & 0.41-0.45 & 0.02-0.03 & 18-25 & $-0.57{\scriptstyle \pm 0.13}$ & $-0.56{\scriptstyle \pm 0.07}$ & $-1.04{\scriptstyle \pm 0.33}$ & $-0.93{\scriptstyle \pm 0.02}$ & $2.22{\scriptstyle \pm 0.25}$ & $1.86$\\	
~ & (\,$x-L$\,) & ~ & ~ & ~ & ~ & ~ & ~ & ~ & ~ & ~ & ~ & ~\vspace{2pt}\\	
\hline 
\hline 
\end{tabular} 
\end{center}
Probed parameter spaces: $0.18\le\beta\le0.65$; $0.30\le b\le0.55$; $0.02\le\kappa\le0.10$; $8\le T\le 40$~K \\
\smallskip 
\end{table*}

Inclusion of data from dust-continuum studies in the analysis of the clump mass-size ($m-r$) diagrams confirms the results from molecular-line mappings (Sect. \ref{mr_diagrams}): consistency between the slopes within 2$\sigma$ and, occasionally, a shift by a factor of up to 3 on the size axis. Generally, the agreement between the locations of the `average clump ensemble' and the observed clumps is better; especially, for the combined clump sample from M\,17 and NGC\,7538. The discrepancies on the $m-r$ diagrams could be attributed to the various clump-finding techniques used in the referred observational studies (cf. Table \ref{table_mc}, column 10) and hence to different clump size definitions. The latter would not lead to substantially different mass estimates if a typical clump structure is a dense  central region and diffuse outer shell. A careful comparative analysis of the basic clump-finding algorithms, applied to identical observational or numerical datasets, is necessary for a more comparison 
between our statistical clump description and the observed clumps in MC complexes.

The role of the magnetic fields in shaping the physical characteristics of MCs and of their substructures can be accounted for more thoroughly in a further extension of this work. \citet{Molina_ea_12} proposed an analytical model of the relation between the width of the PDF $\sigma$ in magnetized turbulent medium. It includes a modification of equation \ref{eq_sigma_PDF}, consistent with a scaling law of magnetic field $B\propto\langle \rho \rangle^{1/2}$ (equation \ref{eq_mag_scaling}) and can be easily incorporated in our modeling.

Finally we caution the reader that the predictions of our model hold for less evolved clumps of sizes $\gtrsim 0.1-0.2$~pc and must not be compared with observational mass distributions of dense (prestellar) cores. The latter ones will be subject of another study which considers the high-density power-law tail of the density PDF \citep[cf.][]{Kainu_ea_09, KNW_11} in active star forming regions as a basis for cores' description. The essence of this approach is presented in \citet{DSV_12} while the model is in process of development. Such study may answer the question whether the difference between the clump mass function and the core mass function is physical or statistical.

\section{Summary}    \label{Summary}
Our statistical approach for physical description of condensations (clumps) formed through a turbulent cascade during the early MC evolution predicts: i) the cloud structure in terms of effective size-mass scaling relations, through the mass-density exponent $x$ as a function of the spatial scale $L$ (DVK11); and ii) the composite clump mass function (ClMF; Paper I). Different models within this framework are generated by choosing an appropriate energy equipartition relation and a set of 4 free parameters: velocity scaling index $\beta$, turbulent forcing parameter $b$, mapping resolution parameter $\kappa$ and temperature $T$. In this Paper we compared ClMFs from molecular-line and dust-continuum studies of Galactic cloud complexes with ones derived from our model applying alternative fitting criteria: fitting the structure $x(L)$ of individual complexes (when additional dust-extinction data are available) or direct fitting of the observational ClMF.  

Both fitting criteria lead to modeled clump mass distributions, in good or excellent agreement with the considered observational data. The equipartition relations which yield these fits indicate gravitational boundedness of the dominant clump population in the considered clouds, possibly including the contribution of magnetic or thermal energy: model cases \wkinii, \wkinth~or \wkinmag. On the other hand, the results for Taurus, M\,17 and NGC\,7538 rather hint at strongly gravitating or contracting inner parts of those complexes (model case \wkiniv). The derived best-fit values of the parameters $\beta$, $b$ and $T$ for all studied individual clouds span relatively narrow ranges. In most MC complexes the typical velocity scaling index is found to be similar to the original ``Larson's first law'' (Eq.~\ref{eq_Larson_1}, $\beta=0.38$) or to $\beta\sim 0.43$ testified from recent observations \citep{Pad_ea_06, Pad_ea_09}. The best-fit values of the turbulent forcing parameter concentrate around $b\simeq0.40$ which corresponds to a natural mixture between compressive and solenoidal modes \citep{Fed_ea_10}.

The modeled clump mass distributions support a ClMF which might be represented as a combination of two-power law functions. The latter are separated by a characteristic mass $M_{\rm ch}$ that varies typically within one order of magnitude or more: from about ten to hundreds solar masses. The slope of the intermediate-mass ClMF is shallow and nearly constant: $-0.25\gtrsim\Gamma_{\rm IM}\gtrsim-0.55$. The high-mass part of the ClMF corresponds to gravitationally unstable clumps in all considered model cases and hence a more appropriate description should take into account the dynamical clump evolution. Such description is achieved through time-weighting of the clump mass distribution. (The effect will be similar if the ClMF is derived from a statistically significant sample containing clumps in MC complexes at different evolutionary stages of clump formation.) We obtained slopes within a broader range $-0.9\gtrsim\Gamma_{\rm HM}\gtrsim-1.6$ that includes the typical value for fractal clouds $-1$ \citep{Elme_97} as well that of the stellar initial mass function \citep{Sal55}.

Comparison between the observational and the modeled clump mass-size diagrams reveals agreement of the slopes within the $2\sigma$ limit and, in most cases, a systematic shift toward smaller or larger clump sizes. The latter is to be attributed to variety of size definitions used in the different clump-finding techniques.

{\it Acknowledgement:} We are grateful to the anonymous referee for his/her insightful comments and recommendations which helped us to improve substantially this Paper. We thank J. Di Francesco for the list of clump properties in Rosette MC derived from {\sc Herschel} data, and N. Schneider for providing us the mass/spatial scale values shown in Fig. 8 from the same Herschel column density map. We appreciate the help of V. Ossenkopf and R. Simon who provided for us data on the individual clump characteristics in Orion B from the original study of \citet{Kramer_ea_98}. 

T.V. acknowledges support by the {\em Deutsche Forschungsgemeinschaft} (DFG) under grant KL 1358/15-1.  


\label{lastpage}
\end{document}